\documentclass[12pt]{article}
\usepackage{epsfig}

\begin{document}

\titlepage
\begin{center}
{\bf \Large Is the observable Universe generic?
\footnote{Proceedings of the IV Mexican School on Gravitation and 
Mathematical Physics, DGM-SMF, 2001. ``Membranes 2000.'' }
}
\end{center}

\vskip 0.6cm

\begin{center}
{\large C\'esar A. Terrero-Escalante 
\footnote{cterrero@fis.cinvestav.mx}
        and Alberto A. Garc\'{\i}a 
\footnote{aagarcia@fis.cinvestav.mx}}
\end{center}
\vskip 0.2cm
\begin{center}
{\small 
\it Departamento de F\'{\i}sica,
Centro de Investigaci\'on y de Estudios Avanzados del IPN,
Apdo.~Postal 14-740, 07000, M\'exico D.F., M\'exico.}
\end{center}

\vskip 0.65cm
\begin{center}
{\large Abstract}
\end{center}
\noindent
Recently an inflationary potential yielding power spectra characterized by a 
scale-invariant tensorial spectral index and a scale-dependent scalar spectral 
index was introduced. We analyze here the implications that this potential 
could have for the large-scale structure formation in the multiverses scenario
of eternal inflation.
\vskip 0.35cm
\noindent

\section{Introduction}
\label{intro}

The Standard Cosmological Model (SCM) and Standard Particle Model (SPM) provide
a framework for successfully describing the evolution of the microworld from
the primordial plasma. Nevertheless, in this framework there are no answers 
to the questions of where that hot soup came from and how were formed such 
large structures like galaxies clusters and galaxies which, in turn, are
necessary to explain
the formation of stars, heavy chemical elements and life. To answer these
questions it is necessary to go beyond the standard knowledge.  

New results of the cosmic microwave background (CMB) observations 
\cite{CMBdata} confirm the inflationary
scenario plus gravitational instability as the leading 
candidate for a theory 
of large-scale structure formation at our Universe \cite{inflation}. 
According to this theory, the fluctuations of matter and space-time 
 that took place during a period of rapidly accelerated
expansion in the early Universe \cite{perturbations}
seeded the formation of large-scale structure and gave rise to anisotropies in
the CMB radiation. The simplest scenario describes this  expansion 
driven by the potential energy of a single real scalar field coined inflaton. 
The primordial fluctuations are characterized in terms of 
power spectra of scalar and tensorial perturbations. Different theories for 
the 
physics of the early Universe lead to different primordial spectra (different
 initial conditions for density evolution) and, after evolving for a while, to
 different distributions of
density and temperature anisotropies in the currently observable Universe. 
Present level of observations accuracy allows a large number of potentials 
being reliable candidates for the inflationary potential. However, near future
measurements to be carried out by satellites Map and Planck \cite{future} 
promise to reduce this number to a few candidates. Then, it could be possible
to obtain some hints about the physics on energies close to the Planck scale
if, as we assume throughout this paper, the single scalar field scenario is
the dominant feature of the early Universe dynamics.

If several extensions of the SPM can reproduce its features in a equally
satisfactory fashion, then the answer to whether our Universe is generic it is 
related to the answer of how probable is a given SPM extension to be picked 
out amongst the others. But, even if one can devise a deterministic way of 
choosing the appropriate SPM extension, different initial values for the 
inflaton field dynamics will lead to different universes. We shall come back to
 this point in the next section.
  
A very plausible outcome of the satellites measurements will be a central 
constant value for 
the rate between the tensorial and scalar perturbations amplitudes as well as
the possibility of fitting the CMB observations using a scale-dependent scalar
index.
Recently, a model generating perturbations described by a 
scale-invariant tensorial spectral index was introduced \cite{ConstNt}. This 
model 
could be the best fit to near future observations if the scalar index can be 
fairly approximated by a constant at small angular 
scales and departs from that behavior at large scales. Taking into account
that currently a constant scalar index is typically used to fit the CMB 
observations, these assumptions for its possible scale-dependence are well
based. Thus, the physics corresponding to this potential could be very close 
to the actual early Universe physics.
 
In this contribution we assume that the above mentioned inflationary potential
 has a strong 
similarity with the actual one and analyze the implications that this 
similarity
would have for the formation of structure in the multiple Universes arising
through the mechanism of eternal inflation. 

\section{Future-eternal inflation}
\label{eternal}

During inflation it is suitable to assume the space-time geometry 
corresponding to 
a flat Friedmann-Robertson-Walker universe. Then the inflaton classical
equations of motions are given by
\begin{eqnarray}
H^2 &=& \frac{\kappa}{3}\left(\frac{\dot{\phi}^2}{2} + V(\phi)\right),
\label{eq:Friedmann} \\
\ddot{\phi} &+& 3H\dot{\phi} = -V^\prime(\phi) \, ,\label{eq:mass}
\end{eqnarray}
where $\phi$ is the inflaton, $V(\phi)$ the inflationary
potential, $H=\dot{a}/a$ the Hubble parameter, $a$ the scale
factor, dot and prime stand for derivatives with respect to cosmic
time and $\phi$ respectively, $\kappa = 8\pi/m_{\rm Pl}^2$ is the
Einstein constant and $m_{\rm Pl}$ the Planck mass. Note that for a unique
solution of this system it is necessary to set the initial conditions as for
$\phi$ as for $\dot{\phi}$.

Combining Eqs.~(\ref{eq:Friedmann}) and (\ref{eq:mass}) it is obtained that
for inflation to proceed, i.e. for $\ddot{a}>0$, we must have $V>\dot{\phi}^2$.
Because friction in Eq.~(\ref{eq:mass}) is almost constant then, for
the inflationary period to be long enough, the potential must be rather flat.
Finally, inflation ends when the inflaton reaches the true vacuum 
(or, as we shall see later, reaches a critical value which triggers the decay 
of 
secondary scalar fields) so, inflation must start far enough from the value
corresponding to the potential absolute minimum 
(or from the triggering critical value). 
That
condition seems to impose some tunning of the initial conditions for inflation
that it is the kind of problem that the inflationary idea is aimed to solve.
This paradox can be resolved if we assume a random initial spatial distribution
for the scalar field values. All we need is a finite probability for a tiny 
patch of the early Universe to have the desired inflaton value. To understand
this, we have to consider the scalar field quantum fluctuations. Indeed, 
being the inflaton a quantum object, the time evolution of the average value
of $\phi$ in a region of linear size $d_H=H^{-1}$ can be described by,
\begin{equation}
\phi(t + \Delta t) = \phi(t) + \Delta_{cl}\phi(\Delta t) 
+ \delta_{qu}\phi(\Delta t) \, ,
\label{InfDyn}
\end{equation} 
where $\phi(t+\Delta t)$ and $\phi(t)$ are the inflaton values 
at times $t+\Delta t$ and $t$, 
$\Delta_{cl}\phi(\Delta t)\simeq \dot{\phi}H^{-1}$ 
and 
$\delta_{qu}\phi(\Delta t)$ are changes of the inflaton value during a Hubble 
time, $\Delta t=H^{-1}$, due to the classical motion and quantum fluctuations
respectively. The random quantum jump $\delta_{qu}\phi(\Delta t)$ is 
characterized by a Gaussian distribution with zero mean and standard deviation 
$\Delta_{qu}\phi\simeq (2\pi)^{-1}H$. Taking into account that 
$V>\dot{\phi}^2$, then, to leading order, $V\simeq H^2$ and
larger $\left|\phi\right|$ values will allow larger deviations of $\phi$ from 
the 
average value. If we assume that in some Hubble regions the inflaton can
pick an initial value $\phi^*$ such that, 
\[
{\Delta_{qu}\phi(\phi^*) \over \Delta_{cl}\phi(\phi^*)} \simeq
\frac1{2\pi}\frac{H^2}{\dot{\phi}} \geq 1 \, ,
\]
then we have a finite
probability (of about a 16\%) for the negative (positive) quantum contribution to
be larger than the positive (negative) classical displacement. The conclusion
is that there is a finite probability that in some Hubble regions of the early
Universe the overall scalar field motion will be in the direction of larger 
$\left|\phi\right|$ values, i.e., higher energies. During inflation, larger
values of $\left|\phi\right|$ correspond to a larger factor
 $e^N$ ($N$ is known as the efolds number) by which the size of the initial 
patch is increased. 
Then, regions 
where the scalar field rolls upward the potential hill inflate more and after
some Hubble times we will have large spatial regions where all the above story
can start again.    
In the remaining regions,
where the overall scalar field motion will be in the direction of smaller 
$\left|\phi\right|$ values, i.e., lower energies, the inflaton eventually will
 reach the potential minimum (or the triggering critical value) and will give
 rise 
to ``big bangs'' leading to different universes. This way, the general picture 
involves some regions of the ``mother'' Universe eternally inflating and a 
large number of ``baby'' universes springing to light through ``big bang'' 
labor, with their large-scale structures seeded by the inflaton and metrics 
quantum fluctuations and growing by gravitational instability. 
In very few words, that is 
what it is known as (future) eternal inflation \cite{eternal}.

Now, for each one of these ``baby'' universes the inflaton final roll-down 
starts at different values of $\phi$ and with different values
of $\dot{\phi}$ and, according with Eqs.~(\ref{eq:Friedmann}) and 
(\ref{eq:mass}), it means that the inflationary period will set different
initial conditions for the ``baby'' universes births. Then, there will be
universes with different degree of flatness, and/or different degree of 
homogeneity and isotropy, and/or different rate of expansion, and/or different
 number of
topological defects, and/or different number and spectrum of particles, and/or
different
large scale-distribution of matter. Summarizing, from the observational point 
of view (if there is life ``there'' to make the observations), most of these 
universes are different each from the other. Thus, the question in our title 
reduces to which is the 
probability for a universe like our to exist in the multiverses scenario. One
can naively conclude that with an infinite number of births of ``baby'' 
universes
this probability must be high, but it is not so simple. The point is that, in
the definition
\[
Probability = {number \; of \; universes \; like \; our \over
total \; number \; of \; universes} \, ,
\]
we are dealing with infinite quantities, $\infty/\infty$, i.e., is an 
ill-defined probability.

\section{The model}
\label{model}

From measurements of CMB polarization to be done by satellite Planck it
could be possible to determine a central constant value for the rate between
tensorial and scalar inflationary amplitudes. From this value it is 
possible
to estimate a constant leading order value for the tensorial index \cite{KK}. 
On the
other hand, the increased range and resolution of the CMB observations must
allow to discern a possible scale-dependence for the scalar index. Taking into
 account that current use of a constant scalar index, corresponding to a
power-law spectrum, yields nice results while fitting the CMB 
spectrum, it must be expected the scale-dependence of the scalar index to be
slender or hidden in the scales out of reach for observations. A potential
yielding perturbations with these peculiarities was found in 
Ref.~\cite{ConstNt}:
\begin{equation}
V(\phi)= \left\{
\begin{array}{rcl}
\phi(\epsilon_1)&=&-\frac{2(C+1)}{\sqrt{\kappa}}\frac{1}{\sqrt{1-4\delta}}
\left[-\sqrt{1+\sqrt{1-4\delta}}\arctan\left(\frac{\sqrt{2\epsilon_1}}
{\sqrt{1+\sqrt{1-4\delta}}}\right)\right.\\
&& + \left. \frac{1}{2} \sqrt{-1+\sqrt{1-4\delta}}
\ln\left|\frac{\sqrt{2\epsilon_1}+\sqrt{-1+\sqrt{1-4\delta}}}
{\sqrt{2\epsilon_1}-\sqrt{-1+\sqrt{1-4\delta}}}\right|\right]
+ \phi_0 \, ,  \\
&& \\
V(\epsilon_1)&=&V_0(3-\epsilon_1)
\left|\epsilon_1 ^{2}+\epsilon_1 +\delta\right|^{C+1}
\left|\frac{2\epsilon_1+1+\sqrt{1-4\delta}}
{2\epsilon_1+1-\sqrt{1-4\delta}}\right|^\frac{C+1}{\sqrt{1-4\delta}}
\, ,
\end{array}
\right.
\label{eq:Sol4}
\end{equation}
where $V_0$ and $\phi_0$ are integration constants, 
$C \equiv \gamma_{\rm E} + \ln 2 - 2\approx -0.7296$,
$\delta\equiv n_T/2$, and $n_T=const.<0$ is the tensorial spectral index. The 
parameter 
$\epsilon_1$ defined by \cite{HFFampl, inflation},
\begin{equation}
\epsilon_1 \equiv {{\rm d} \ln d_{\rm H}\over {\rm d} N} =
 3 { \frac{{\dot{\phi}}^2}{2} \over \frac{{\dot{\phi}}^2}{2} + V(\phi)} \, ,
\label{eq:HFF1}
\end{equation}
measures the logarithmic change of the Hubble distance per e-fold of 
expansion and during inflation is restricted to the interval 
$\epsilon_1\in[0,1)$. In fact, in 
Ref.~\cite{ConstNt} it was shown that this potential is actually inflationary
in only two
\begin{figure}[ht]
\centerline{\psfig{file=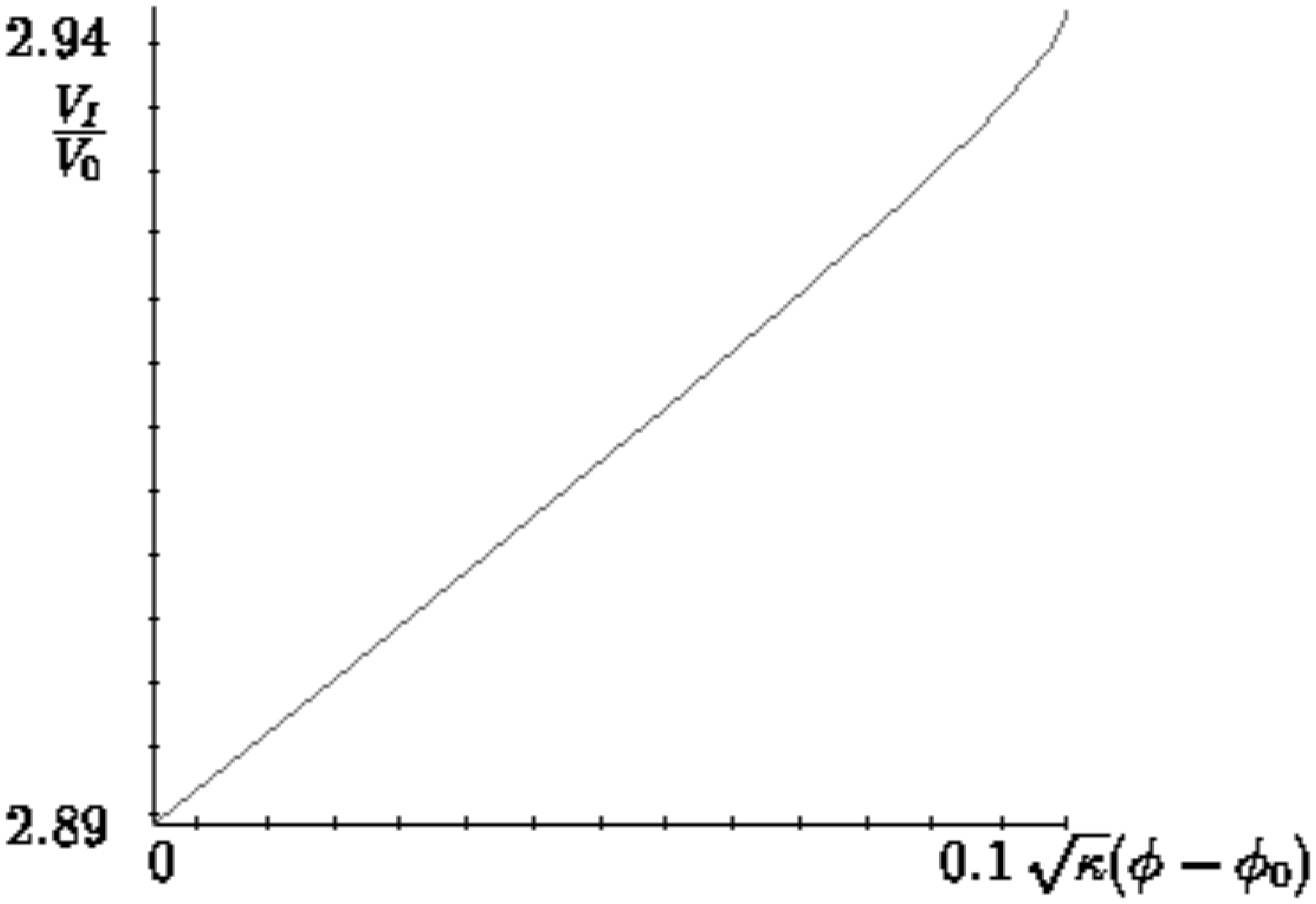,width=7.cm} 
\psfig{file=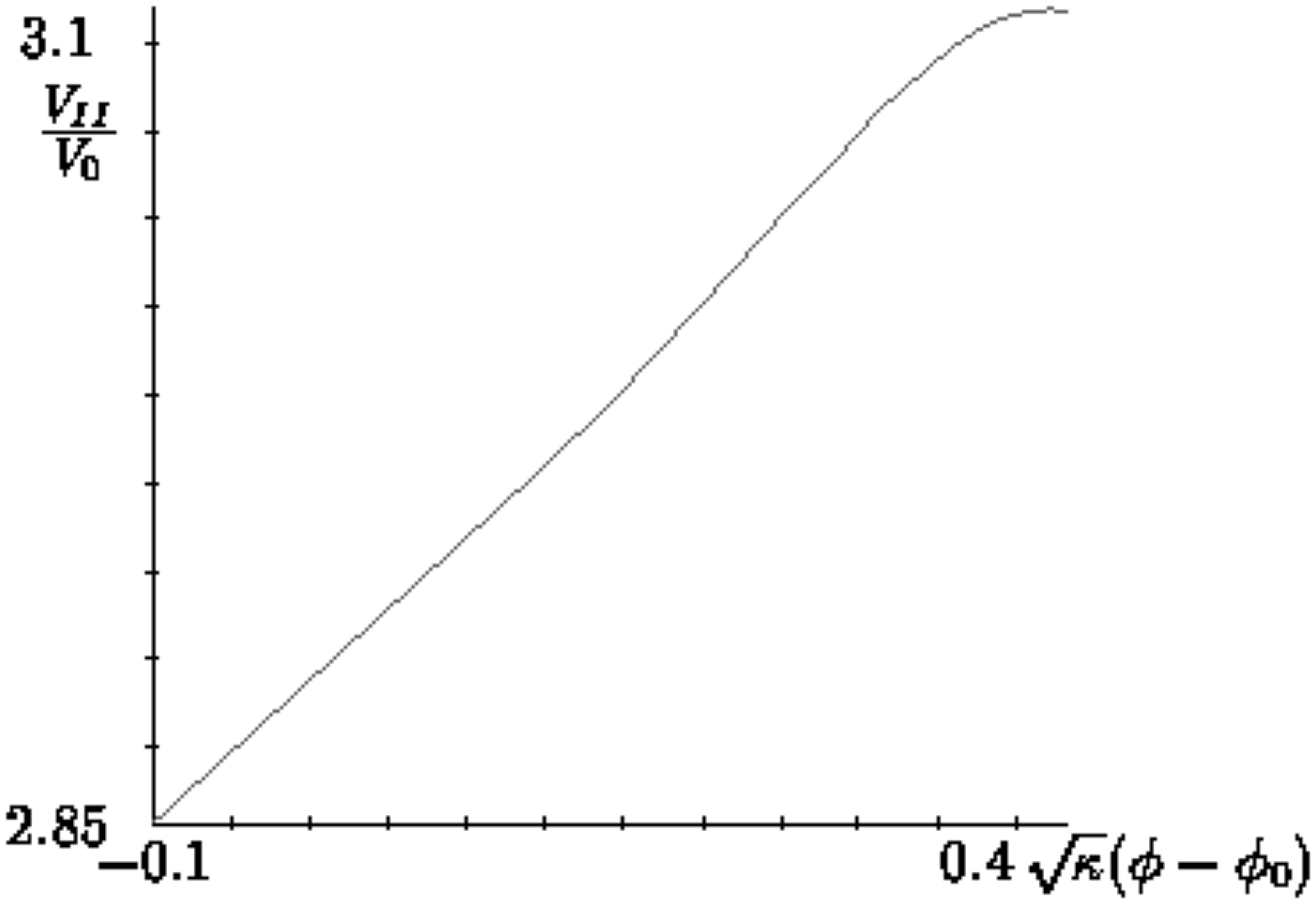,width=7.cm}}
\caption{Plots of the potential branches, $V_I$ (left) and $V_{II}$ (right).}
\label{fig:VI_II}
\end{figure}
branches corresponding to two different intervals of values for 
$\epsilon_1$, namely, 
$I=[0,\epsilon_1^+)$ and $II=(\epsilon_1^+, \epsilon_1^*]$,
where $\epsilon_1^+$ and $\epsilon_1^*$ are some constant values. A plot of
these branches is presented in Fig.~\ref{fig:VI_II},
where a value of $\delta=-0.01$ was used. In this figure the potential is
scaled by $V_0^{-1}$ and the scalar field by $\sqrt{\kappa}$.
Let us note that, for $V_I$ and $V_{II}$, 
$\epsilon_1(t)\rightarrow\epsilon_1^+$ 
is reflected in the inflaton asymptotically rolling down toward the origin. 

For this model the amplitudes of the scalar perturbations as a function of 
scales can be also given parametrically. The scales depend on $\epsilon_1$ as,
\begin{equation}
k= k_0\left|\epsilon_1 ^{2}+\epsilon_1
+\delta\right|^\frac{C+1}{2}
\left|\frac{2\epsilon_1+1+\sqrt{1-4\delta}}
{2\epsilon_1+1-\sqrt{1-4\delta}}\right|^{\frac{3(C+1)}{2\sqrt{1-4\delta}}}
\,,
\label{eq:k3}
\end{equation} 
where $k_0$ is an integration constant. In turn, the scalar amplitudes 
normalized by the integration constant $A_0$ are given by,
\begin{eqnarray}
\label{eq:dlt0Ae}
\ln \frac{A_S(\epsilon_1 )^2}{A_0^2} 
&=&\frac{\delta \,C-1-C}{C+1}\ln \epsilon_1
+\delta \left( C+1\right) \ln \left| \epsilon_1 ^2+\epsilon_1 +\delta \right|
\nonumber \\
&&+\ \frac{3\delta \left( C+1\right) }{\sqrt{1-4\,\delta }}\ln \left| \frac{%
2\,\epsilon_1 +1+\sqrt{1-4\,\delta }}{2\,\epsilon_1 +1-\sqrt{1-4\,\delta }}%
\right| \nonumber \\
&&+\ \frac{\epsilon_1 ^3+2C\epsilon_1 ^2+2C\delta }{2\left( C+1\right)
\epsilon_1 \,}  \, .
\end{eqnarray}
\begin{figure}[ht]
\centerline{\psfig{file=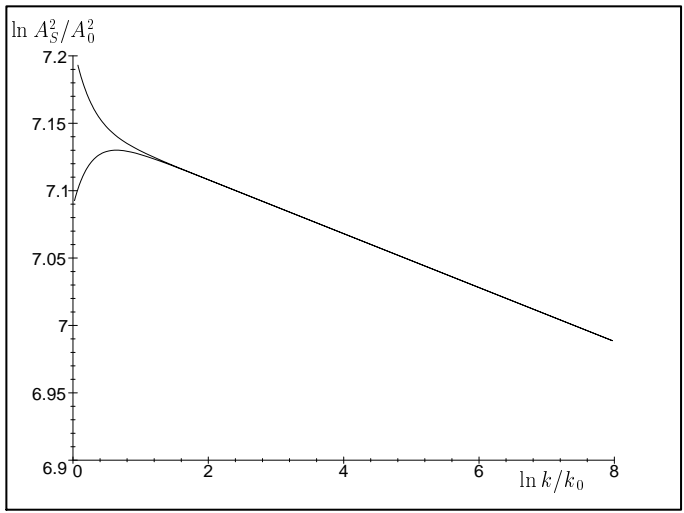,width=9.5cm}}
\caption{$\ln A^2_S$ as a function of the $\ln k$ for $\delta=-0.01$
and for $\epsilon_1\in I$ (lower branch) 
and $\epsilon_1\in II$ (upper branch).}
\label{fig:dlt0lnAlnk}
\end{figure}
As observed in Fig.~\ref{fig:dlt0lnAlnk}, where the parametric
plots of scalar amplitudes for $\epsilon_1\in I$ 
and $\epsilon_1\in II$ is presented in the same figure, differences 
are almost
impossible to note when the full range of scales is considered. The strong 
similarity at
small angular scales (large $k^{-1}$) is given by the asymptotic behavior
$\epsilon_1\rightarrow\epsilon_1^+$. At these scales, both
spectra are very similar to a power-law.
Differences from power-law arise at large angular scales (small $k^{-1}$) 
which could be out of reach for measurements. 
Now, could this potential be related to any physics?
First, our model lacks a graceful exit into the SCM, i.e., there is no
``natural'' way out of inflation here. To solve this problem,  
$\phi$ can be regarded as the dominant scalar field in
a hybrid scenario with $\epsilon_1^+$ being the value corresponding
to the critical value of $\phi$ near which the false
vacuum becomes unstable and the multiple scalar fields roll to the
true potential minimum \cite{inflation,hybrid}.
Next, we note that we cannot make any statement about 
the potential beyond the intervals 
$I$ and $II$
\cite{ConstNt}. 
Then, we can only try to check if the forms of the branches $V_I$ and $V_{II}$ 
in Fig.~\ref{fig:VI_II} arise in any known physics. 
\begin{figure}[ht]
\centerline{\psfig{file=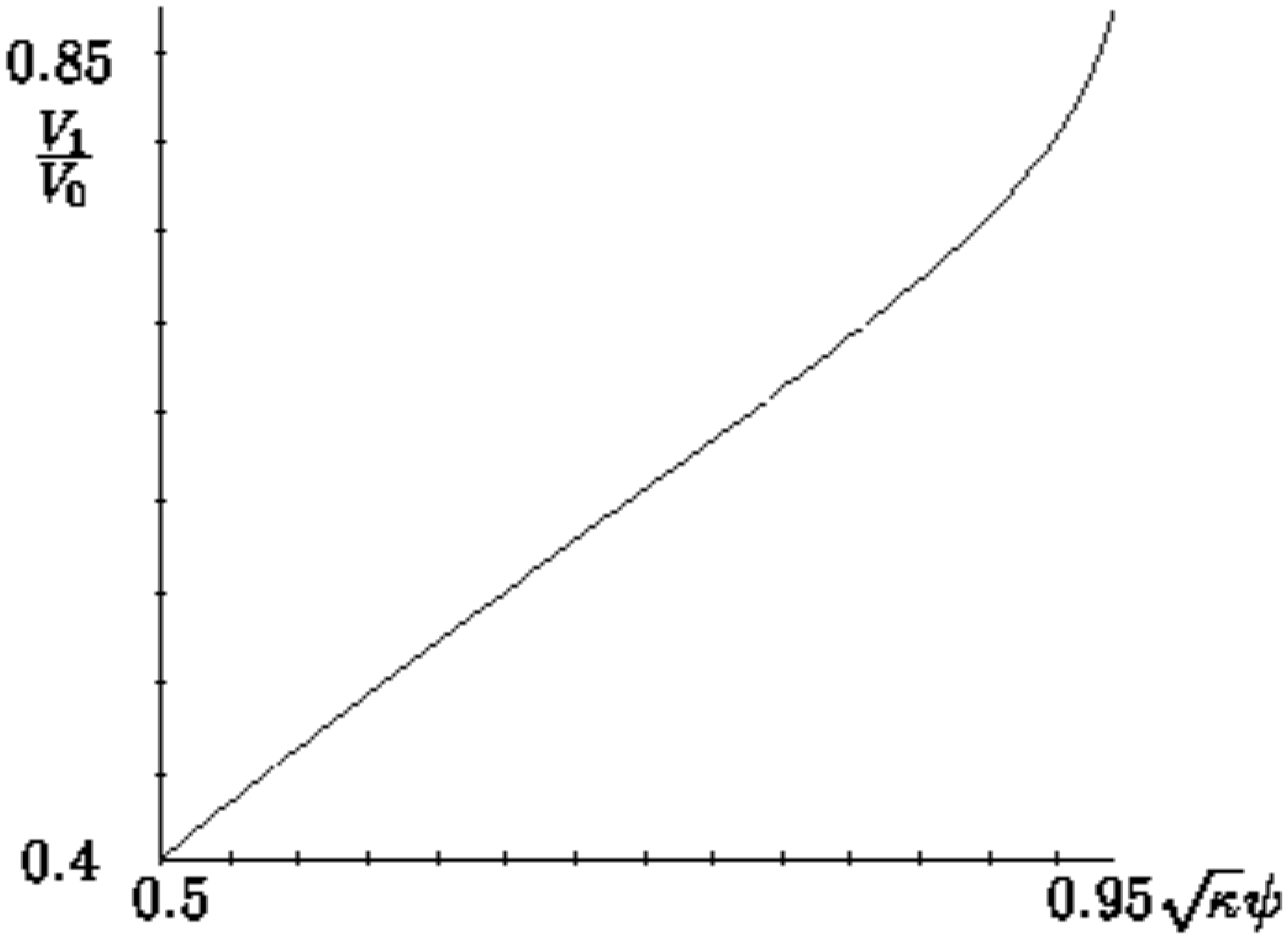,width=7.cm} 
\psfig{file=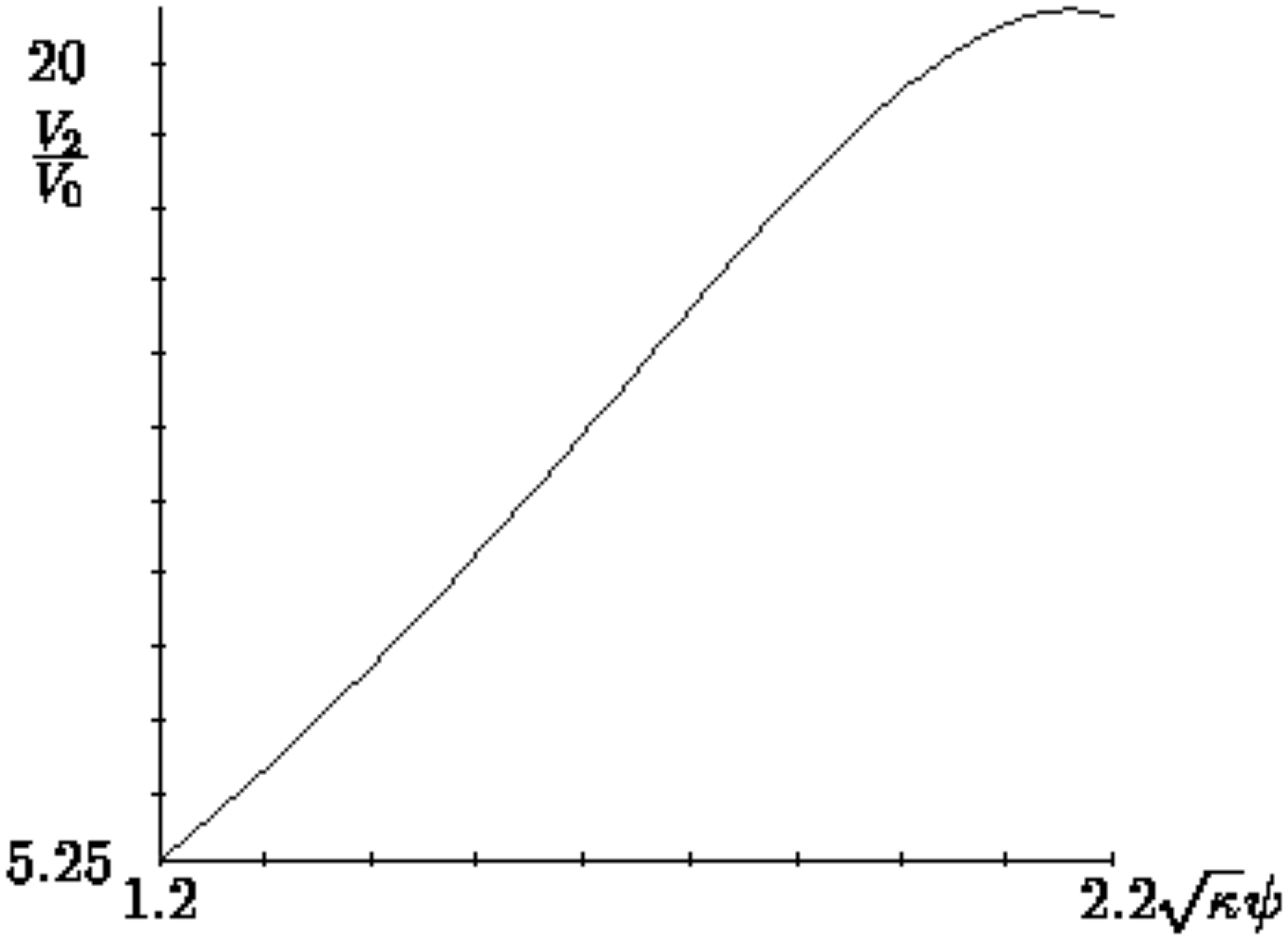,width=7.cm}}
\centerline{\psfig{file=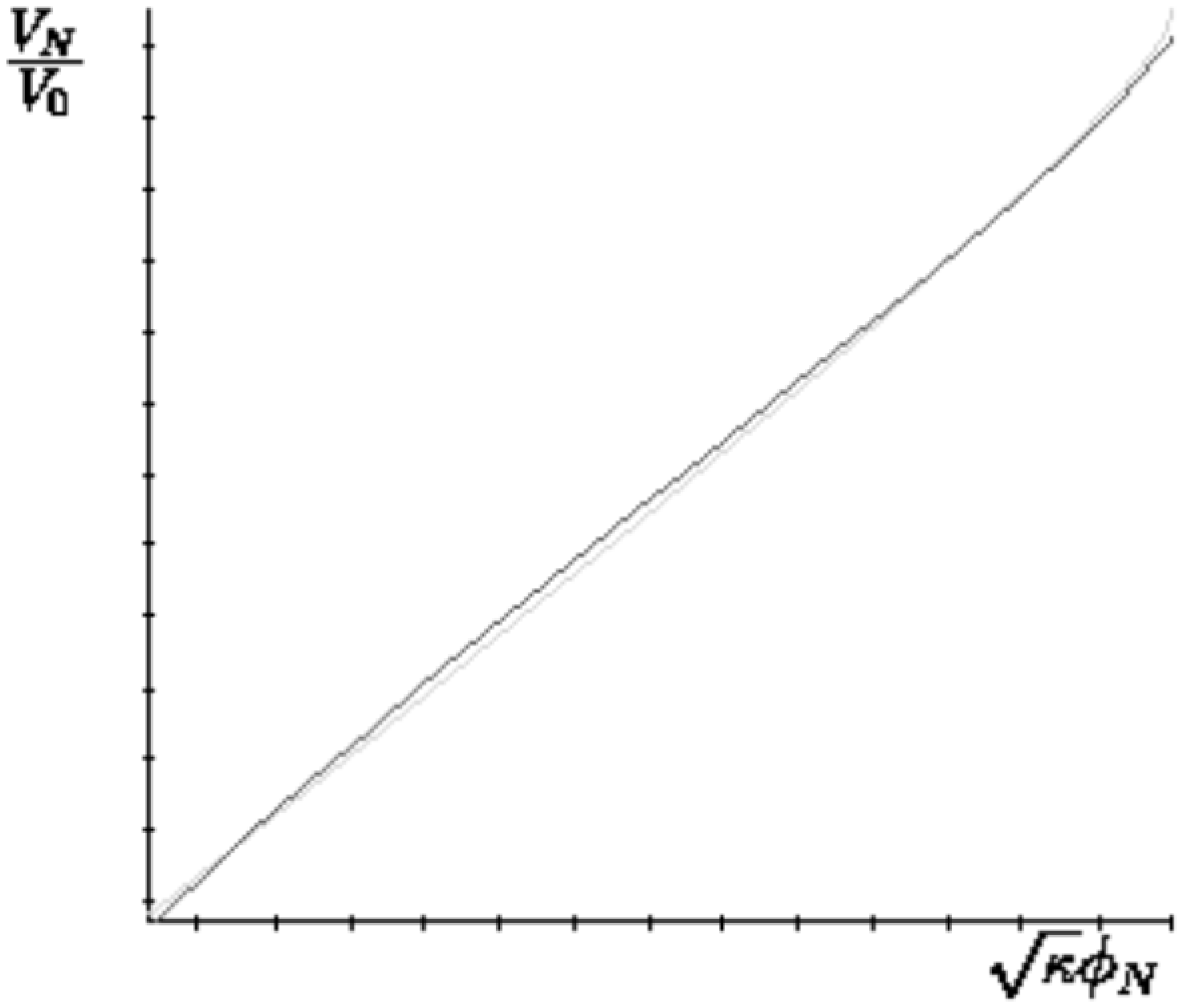,width=6.cm} 
\psfig{file=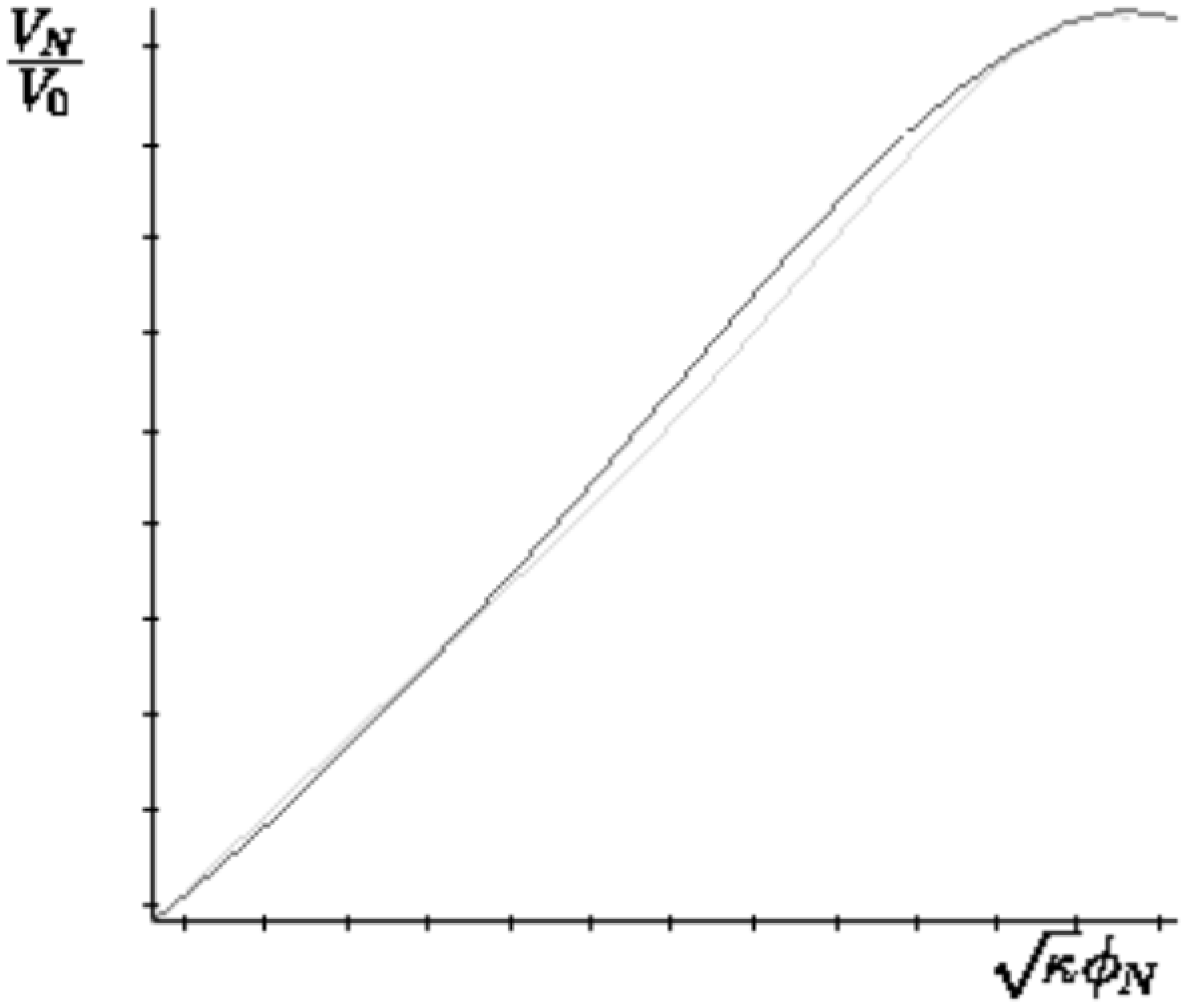,width=6.cm}}
\caption{Plots of potentials $V_1$ (top, left) and $V_2$ (top, right). 
Bottom: plots superposition; 
left, $V_I$ and $V_1$; right, $V_{II}$ and $V_2$.}
\label{fig:V1_2}
\end{figure}
In Fig.~\ref{fig:V1_2} we plotted the potentials $V_1$ and $V_2$ given by,
\[
\frac{V_1}{V_0}= 1 + e^{1.1\psi}\left(1.1\psi^2 
- 0.1\psi^4 - 0.01\psi^6\right) + 0.001\ln \psi \, ,
\]
and
\[
\frac{V_2}{V_0}= 1 - \frac{1}{2} e^{0.7(1-\psi)}
\left(1 - \psi\right)^{\frac{1}{4}} \, ,
\]
at given ranges of the scalar field $\psi$. These potentials resemble
a hybrid inflation model ($V_1$) and a mutated hybrid inflation model ($V_2$)
which have well motivated particle physics behind. The new feature here are
the factors $\exp{\alpha_i\phi}$ (with $\alpha_i$ some constants) that can be
regarded like running couplings (see report 
\cite{PartPhysInf} for details and references about similar models). In the
bottom of 
Fig.~\ref{fig:V1_2}, superposition of $V_I$ with $V_1$, and $V_{II}$ with 
$V_2$ are plotted, after 
conveniently renormalizing the potentials and scaling and shifting the scalar 
fields. It can be observed that an 
opportunity window is open for potential (\ref{eq:Sol4}) being linked
to an (exotic) extension of the SPM. We use this admittedly ad hoc argument
just to note that physics described by potential (\ref{eq:Sol4}) deserve
more attention than if it was merely a nice tool to fit the observational 
data.

\section{Generic or not?}

There is no reason to assume a stochastic initial distribution
for $\phi$ and not for $\dot{\phi}$. According to 
Eq.~(\ref{eq:HFF1}), starting with random $\phi$ and 
$\dot{\phi}$ means starting with random $\epsilon_1$ then,
in the beginning, in some Hubble regions the inflaton is located in branch 
$V_I$, 
and in some other Hubble regions, in branch $V_{II}$. Hence, in the subset of
these regions where the roll-up the potential will eventually dominate, 
different high
energies physics will be set up. Now, while rolling down, $\Delta_{qu}\phi$
decreases and, due to the uncertainty principle, the equal-time standard 
deviation for 
the canonical conjugate of $\phi$, i.e., $\Delta_{qu}\dot{\phi}$ must increase.
Taking into account 
the asymptotic behavior at these energies, 
$\epsilon_1\rightarrow\epsilon_1^+$, any quantum 
perturbation of the
canonical momentum $\dot{\phi}$ (translated into quantum perturbation of
$\epsilon_1$) makes possible that the inflaton ``jumps'' from branch $V_I$
to branch $V_{II}$ and vice versa. This way, memories from the 
corresponding
high energies physics will be smoothly erased. Moreover, this asymptotic 
behavior of $\epsilon_1$ ensures that most of the quantum fluctuations were
produced close to the end of inflation. These are the perturbations which play 
the most important role from the point of view of large-scale
structure formation. As we have already seen, 
at these scales, 
the scalar spectrum produced while rolling down 
$V_I$ or $V_{II}$ are almost identical to a power-law spectrum. 

\section{Conclusions}
\label{sec:Concl}
If the branched potential with scale-invariant tensorial spectral 
index has something to do with reality, then the ``baby'' universes in the
eternal inflation picture will be very similar each to the other from the
observational point of view.
It means that our Universe will be a generic
outcome of the cosmological evolution rather than some extraordinary event.
These conclusions can break down if any of the assumptions behind 
our calculations occurs to be feeble.
In our opinion, the weighty but arguable assumptions here 
are a constant tensorial spectral index, the second order precision of the 
Stewart-Lyth expressions for the spectral indices, and a semi-classical 
analysis close to the Planck era. 
A study is in process to find out to
which extent a change in any of these assumptions changes our conclusions.

\vspace{0.15cm}
{\bf{Acknowledgments.}}
We are grateful to J.~E.~Lidsey, E.~J.~Copeland and R.~Abramo for useful 
discussions. 
This work is supported in part by the CONACyT 
grant 32138--E and the Sistema Nac. de Investigadores (SNI).

\end{document}